Growth-Induced Unconventional Magnetic Anisotropy in Co/Fullerene (C$_{60}$) Bilayer Systems; Insights from a Two-Grain Stoner-Wohlfarth Model

Sonia Kaushik[1], Rakhul Raj[1], Pooja Gupta[2,3], R. Venkatesh[1], Andrei Chumakov[4], Matthias Schwartzkopf[4], V. Raghavendra Reddy[1], Dileep Kumar[1, a)]

[1] UGC-DAE Consortium for Scientific Research, Khandwa Road, Indore-452001, India
[2] Synchrotron Utilization Division, RRCAT, Indore-452013, India
[3] Homi Bhabha National Institute, Training School Complex, Anushakti Nagar, Mumbai 400094, India
[4] Photon Science, DESY, Notkestraße 85, 22607 Hamburg, Germany

a) Corresponding author: dkumar@csr.res.in

## ABSTRACT

Organic spintronics has drawn the interest of the science community due to various applications in spin-valve devices. However, an efficient room-temperature Organic Spin Valve device has not been experimentally realized due to the complicated spin transport at the metal-organic interfaces. The present study focuses on a comprehensive understanding of the interfacial properties essential for advancing device performance and functionality. The structural and magnetic properties of the ultra-thin Cobalt (Co) films deposited on the fullerene (C$_{60}$) layer are studied to investigate the origin of magnetic anisotropy in the metal-organic bilayer structures. Due to the mechanical softness of C$_{60}$, penetration of ferromagnetic Co atoms inside the C$_{60}$ film is confirmed by the X-ray reflectivity and Secondary Ion Mass Spectroscopy measurements. Grazing incidence small-angle X-ray scattering and atomic force microscopy provided information regarding the structural and morphological properties of the Co/C$_{60}$ bilayers, angular dependent Magneto-optic Kerr effect measurements with varying Co layer thickness provided information about the growth-induced uniaxial magnetic anisotropy. In contrast to the inorganic silicon substrates, magnetic anisotropy in Co film tends to develop at 25 Å thickness on the C$_{60}$ layer, which further increases with the thickness of Cobalt. The anomalous behavior in coercivity and remanence variation along the nominal hard axis is explained by a two-grain Stoner-Wohlfarth model with intergranular exchange coupling. It is further confirmed by a non-uniform spatial distribution of magnetic domains investigated through Kerr microscopy. These anomalies could be attributed to the distribution of magneto-crystalline anisotropy and inhomogeneous strain caused by the formation of a diffused layer at the Co/C$_{60}$ interface. These findings shed light on the intricate relationship between ferromagnetic materials and organic semiconductors, opening up possibilities for customizing magnetoresistance effects and deepening the fundamental understanding of organic spintronic devices.

## INTRODUCTION

The study of ferromagnetic-organic material (FM-OM) thin film nanostructures has gained considerable attention due to their unique physical phenomena and potential applications in organic spintronic devices.[1,2] Based on organic semiconductors (OSC), these structures offer advantages such as low spin-orbit coupling, reduced hyperfine interactions, and long spin lifetime. Additionally, OSC-based devices are cost-effective and mechanically flexible, making them promising candidates for future organic spintronics applications.[2,3] The various FM layers (such as FeCoB,[4] Fe,[5–8] Co,[9,10] FeNi[11]) have been combined with organic material layers ($C_{60}$, Alq$_3$, rubrene, pentacene) over the past decade to achieve controlled magnetoresistance and giant spin polarization.[9,12–14] Among these combinations, $C_{60}$-based FM-OM thin film structures are particularly significant due to the potential for significant room temperature magnetoresistance.[15,16] Considerable progress has been made in understanding spin injection, manipulation, and detection in FM-OM thin film structures. However, achieving efficient spin injection remains challenging, and a comprehensive experimental understanding is still lacking. One of the primary reasons for this might be the mechanical softness of organic semiconductors (OSC) and the complex magnetism at the interfaces. Issues such as metal atom penetration, diffusion, and potential chemical reactions at the interfaces contribute to the complications. Extensive efforts are underway to improve the interface quality[17,18], where techniques such as depositing the FM layers at an oblique angle,[19] conducting depositions under partial vacuum conditions, and at low temperatures[20] are being employed to control the diffusion. Researchers are also studying the magnetic properties of the electrodes, including magnetic anisotropy, magnetic dead layer, and coercivity ($H_C$). The goal is to establish correlations between these magnetic properties and the evolving structure and morphology during film growth.

Magnetic anisotropy is one of the important properties in FM-OSC based structures, and it plays a vital role in controlling the performance and operation of these organic spin valves and determines the magnetic functionality of the devices.[21,22] A recent investigation[23,24] shows the in-plane uniaxial magnetic anisotropy (UMA) of ferromagnetic films, where spins are preferably aligned in the film plane, could be obtained due to the diffused interface morphology of metal-organic structures. The origin of such spin alignment in the preferential in-plane direction is surprising in the FM-OM system. In the case of epitaxial magnetic thin films, UMA is often realized when deposited on various single-crystal substrates such as Ag (001), Cu (100), MgO (100), and SrTiO$_3$ (001).[25–27] However, spin-orbit coupling (SOC) originated through ordered crystallographic structure and is responsible for the preferred magnetization direction with respect to the crystal structure and is known as magneto-crystalline anisotropy (MCA). Unlike inorganic layered structures, where the epitaxy is due to the underneath lattice-matched substrates, the growth of metallic thin films on organic material is either amorphous or polycrystalline in nature due to lattice mismatching and diffusion at the interfaces.

Moreover, polycrystalline and amorphous films do not exhibit long-range structural order.[28] Therefore, it is not easy to realize MCA or UMA from randomly oriented grains in such thin films. Contrary to this, the origin of UMA in organic material-based bilayer structures is obtained and related to the preparation conditions and surface roughness. I. Bergenti, *et al.* grew Co thin films by RF sputtering on Tris(8-hydroxyquinoline) aluminium (Alq$_3$) layers. The origin of UMA in Co film is attributed to the morphology of the underneath Alq$_3$ layer.[23] V. Kalappattil *et al.* have investigated the role of magnetic anisotropy of the bottom electrode in LSMO/Alq$_3$/Co spin valve structure in which they have exploited the substrate-induced compressive strain to increase the magnetic anisotropy of the LSMO layer.[29] Besides the vast importance of UMA, little is known about its origin in metal-organic frameworks. Although previous work correlates the morphology of the organic layer with the origin of magnetic anisotropy, a clear picture of how an isotropic surface roughness can induce magnetic anisotropy in the metal-organic system has not been discussed.[24] A careful and systematic investigation is needed to explore the development of magnetic anisotropy and its cause in a metal-organic bilayer structure.

It is to be noted that organic materials such as pentacene, rubrene, or Alq$_3$ have very low density; thus, the spacer layer with these materials has largely diffused ferromagnetic metals when deposited on top of it. $C_{60}$, being a robust material, provides comparatively good resistance to the penetration of top ferromagnetic material. Along with that, zero hyperfine interaction of $C_{60}$ leads to a longer spin diffusion length and larger spin relaxation times, which are an important parameter for transporting spin-polarized charge carriers. Cobalt is a ferromagnetic material with high saturation magnetization and high curie temperature. Thus, it has been used extensively as the top ferromagnetic layer in vertical spin valve devices. The magnetic properties of Cobalt grown on silicon or quartz substrates are well studied in the literature[26,30–33]. The deposition of Cobalt on an organic layer provides critical understanding regarding the interfacial damage that affects its electrical and magnetic properties. Thus, the present work is focused on understanding the UMA with varying thicknesses of the Co films over the $C_{60}$ layer, where the thicknesses of Co layers are decided based on the *in situ* magnetic investigation of Co on $C_{60}$, as reported in the literature.[10] For this purpose, four Co/$C_{60}$ bilayer samples and the reference samples on bare silicon substrates are prepared with varying thicknesses to understand the magnetic behavior and to correlate the UMA with the structure, as well as the morphology evolved during the film growth. In contrast to previous studies lacking thickness dependence, our study examines a series of bilayers with increasing thickness, ensuring consistent morphology within the bilayers due to identical preparation conditions. This approach enables us to investigate the fundamental mechanism underlying the development of UMA in metal-organic thin films as thickness increases. Gaining a comprehensive understanding of UMA in metal-organic thin films and developing effective control will empower researchers to tailor film properties and optimize their functionality.

**EXPERIMENTAL**

A set of four Co/$C_{60}$ bilayers and reference samples were prepared by depositing with Co films of nominal thicknesses 10 Å, 25 Å, 40 Å, and 180 Å on a 1000 Å thick $C_{60}$ layer. Further, in the manuscript, these samples will be denoted as S1, S2, S3, and S4 and the corresponding reference samples, respectively. Metallic layers were prepared using high-power impulse magnetron sputtering (HiPIMS) under a base pressure of $1.33 \times 10^{-7}$ mbar, whereas the $C_{60}$ layer was deposited on a silicon substrate *via* sublimation of $C_{60}$ powder supplied by Sigma Aldrich ® at a rate of 0.2 Å/s using a homemade thermal evaporator under base pressure of $10^{-6}$ mbar. Before the Co film is deposited, the $C_{60}$ films are characterized in detail to illustrate the quality of the film. The thicknesses of the Co layer on the $C_{60}$ layer were decided as per *in situ* transport measurements performed during the growth of the sample.[10] This study revealed the Volmer-Weber growth of Co on $C_{60}$, where small isolated islands grow larger to connect with other islands and eventually combine into a continuous thin film at around 75 Å thickness. This study presents four samples covering different stages of the sample growth; diffused isolated islands (S1-10 Å), bigger islands (S2-20 Å), connecting islands or percolation stage (S3-40 Å), and continuous layer (S4-180 Å). The thin capping layer of Aluminium (Al) having a thickness of 20 Å was deposited to prevent the oxidation of Co film. Magnetic hysteresis loops of the samples were taken using the Magneto-optic Kerr effect (MOKE) in longitudinal geometry. The loops were recorded by varying the in-plane azimuthal angle from 0° to 360° to investigate magnetic anisotropy in the samples. The thickness and structure of the films were obtained using X-ray reflectivity (XRR) carried out using a Bruker © D8 diffractometer using x-rays of energy 8.047 keV. Synchrotron-based X-ray diffraction (XRD) measurements were carried out at BL02 beamline, RRCAT, Indore in out-of-plane (OP-XRD) and in-plane (IP-XRD) geometry at energy 15 keV. Depth profiling measurements were carried out using Secondary Ion Mass Spectroscopy using $O_2^+$ ions to sputter out the film with energy (30 keV, current 30nA) in the sample area of around $100 \times 100$ µm². For the topographical analysis and surface roughness measurement of the samples, atomic force microscopy (AFM) measurement in tapping mode was performed at room temperature using Bruker's Bioscope Resolve system. The silicon cantilever with a nominal spring constant of 50 N/m and resonant frequency around 290 kHz was used for imaging. To understand the morphological information of the surface and the buried layers, GISAXS measurements were carried out simultaneously with grazing-incidence wide-angle X-ray scattering (GIWAXS) at P03 beamline (synchrotron PETRA III, DESY, Hamburg), using photon energy of 12 keV with a beam size of 35 × 25 µm² (horizontal to vertical ratio). The sample-to-detector distance (SDD) was set to 214 mm, and the LAMBDA 9M pixel detector (X-Spectrum ©) with pixel size of 55 × 55 µm² was used for GIWAXS measurements. While PILATUS 2M pixel detector (Dectris AG ©) (172 × 172 µm²) was kept at the SDD of 4280 mm for GISAXS measurements.

## RESULTS AND DISCUSSION

The growth of the $C_{60}$ layer on Silicon substrates was studied and characterized using XRR, AFM, GISAXS, and GIWAXS measurements. The room temperature deposited fullerene film is polycrystalline in nature, and from the XRR measurements, the roughness of the $C_{60}$ film is found to be in the order of 16 Å (not shown here). The AFM micrographs confirm the uniform morphology of the $C_{60}$ film after depositing the $C_{60}$ film through the sublimation of $C_{60}$ powder. Further, Raman Spectra confirms the deposited film is of $C_{60}$ and exhibits all the Raman modes as already discussed in the literature.[10] Figure 1 (a)-(d) shows the X-ray reflectivity (XRR) patterns of all samples, which are plotted against the scattering vector $q_z = 4\pi \sin\theta/\lambda$ on the x-axis, where $\theta$ is the incident angle, and $\lambda$ is the wavelength of the X-ray. The periodic oscillations (Kiessig fringes) in all the XRR patterns are due to the thickness of the total structure.[34] In the case of the S4 sample, broad and short periodic oscillations in the XRR pattern correspond to the Co and total thickness (Co+$C_{60}$), respectively. The difference between two periodic oscillations is inversely proportional to the thickness of the layer ($d = 2\pi/\Delta q$), where $d$ is the film thickness. The broad oscillations are not visible in the case of the ultrathin Co layers. The more damping in the oscillations at the lower thickness of Co (sample S1, S2, and S3) is attributed to the rough interface due to the deep penetration of Co metal atoms into the $C_{60}$ layer.[35]

All XRR patterns are fitted using GenX software to extract the structural parameters of all the samples. The thickness of the $C_{60}$ layer varies from 970 Å to 1000 Å, depending on the amount of diffusion of the Co clusters inside the $C_{60}$ layer. The thickness and density of the Co film are shown in Table 1. Compared to the reference samples, the best fit to the XRR data is obtained by considering two different layers ($Co_{top}$ and $Co_{int}$) of Co film with varying densities. $Co_{int}$ layers have lower $\rho$ in all samples due to Co diffusion in the $C_{60}$ layer near the interface. In the case of S1 and S2 samples, XRR was fitted by considering the whole Co layer as the $Co_{int}$ layer having a density of 5.9 g/cc and 6.95 g/cc, respectively. It is mainly due to the lower deposited thickness of Co film, which gets completely diffused inside the $C_{60}$ layer. In sample S3, the reduced density of the $Co_{top}$ layer (about 86%) compared to bulk Co material ($Co_{bulk}$~8.9 g/cm³) indicates that the $Co_{top}$ layer is still not entirely continuous.

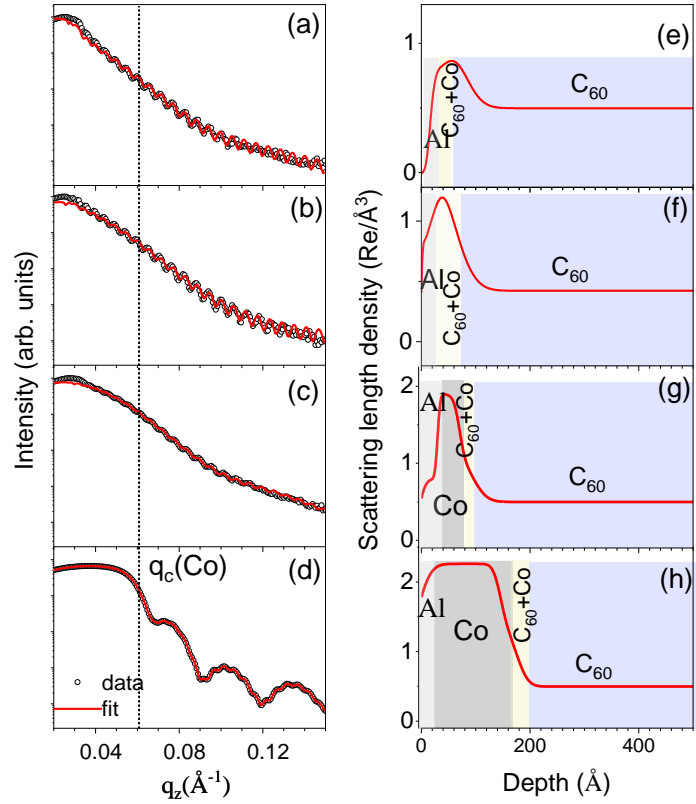

**Figure 1** (a)-(d) X-ray reflectivity pattern, circles represent the experimental data obtained and solid line represents the fitted data and (e)-(h) scattering length density profile of the samples S1, S2, S3 and S4, respectively, where Co, $C_{60}$, and Al are used to represent the Cobalt, fullerene, and Aluminium layer in the samples.

**Table 1** Thickness and density of Co film obtained from XRR fitting. Errors in layer thicknesses are ± 0.5 Å.

|         | $Co_{top}$ |           | $Co_{int}$ |           |
|---------|------------|-----------|------------|-----------|
| Sample  | $d$ (Å)    | $\rho$(g/cc) | $d$ (Å)  | $\rho$ (g/cc) |
| S1      | -          | -         | 17         | 5.9       |
| S2      | -          | -         | 32         | 6.95      |
| S3      | 40         | 7.67      | 21         | 4.9       |
| S4      | 154        | 8.9       | 31         | 5.6       |

On the other hand, the density of the $Co_{top}$ layer for S4 is almost the same as the Co bulk material, confirming the formation of a continuous layer on the $C_{60}$ film. This is per the earlier *in situ* work [10], where Co film grows on $C_{60}$ *via* a diffused small isolated island in the $C_{60}$ matrix at the interface. These isolated islands grow larger to connect with other islands, which combine to form a continuous film at a higher thickness, around 75 Å. The thickness of the $Co_{int}$ increases from 21 Å to 31 Å on increasing the deposition thickness from S3 to S4. The XRR data fit reveals information regarding the diffusion/intermixing of the Co and $C_{60}$ at the interface, which is further confirmed by the elemental depth profiles observed in the Secondary Ion Mass spectroscopy measurements, as shown in Fig. 2(a) and (b) for S4 reference and S4 samples respectively.

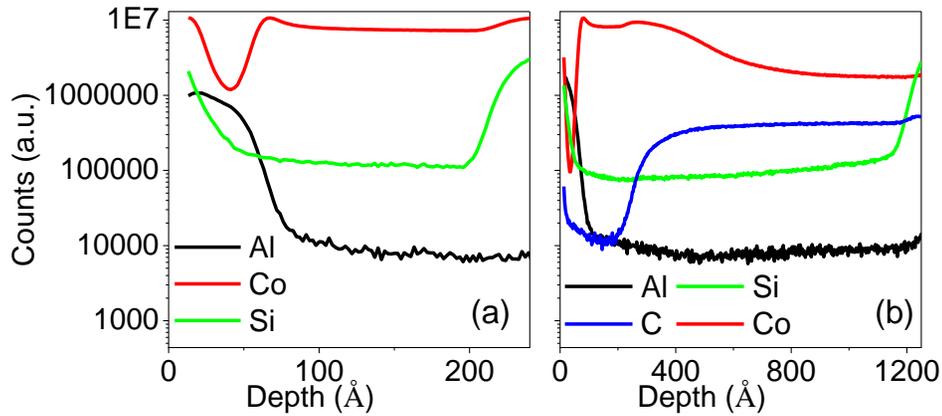

**Figure 2** (a)-(b) SIMS depth profile of Co, Al, C, and Si for sample S4$_{ref}$ and S4, respectively.

The morphological and structural characterization of the samples was done using AFM, GISAXS, and GIWAXS measurements. Figure 3 (a)-(d) gives the typical AFM micrographs for all samples. The average surface roughness values are obtained by extracting the line profiles from different parts of the AFM image. The final r.m.s. roughness values obtained are 22.4 Å, 20.5 Å, 26.1 Å and 22.0 Å for samples S1, S2, S3 and S4, respectively. The maximum roughness is obtained in sample S3. The mean size of Co grains is estimated using the ImageJ software and is obtained as 15 nm, 27 nm, and 36 nm for samples S2, S3, and S4, respectively. Slight elongation of the particles observed along the marked arrow of the samples S3 and S4 images may be due to anisotropy in nucleation and growth or preferential percolation similar to the earlier study.[30]

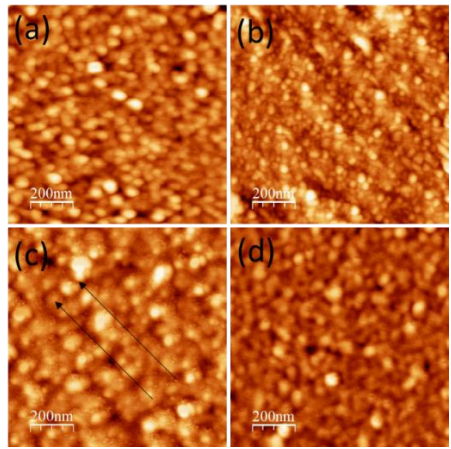

**Figure 3** (a)-(d) shows AFM micrograph for sample S1, S2, S3 and S4 respectively.

GISAXS measurements were carried out simultaneously with GIWAXS. The schematic of the experimental geometry used at the P03 beamline, PETRA III, is shown in Figure 4. Here, $k_i$ and $k_f$ denote the incident and reflected wave vectors $\alpha_i$, and the incidence angle, $2\theta_f$, and $\alpha_f$ are the exit angle and in-plane scattering angles, respectively. In 2-dimensional GISAXS images, horizontal line cuts taken at the Yoneda peak region along $q_y$ direction provide information on the lateral structures of the material parallel to the film surface, while the off-specular vertical cut along $q_z$ offers information on the structure perpendicular to the substrate.[36]

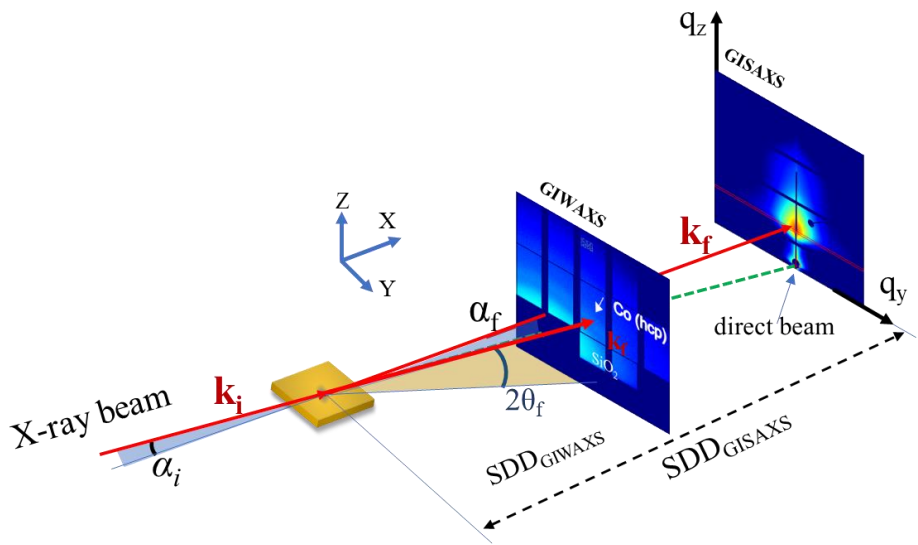

**Figure 4** Schematic of the geometry of the experiment for GISAXS and GIWAXS measurements.

Two-dimensional (2D) GISAXS images of the samples are shown in Fig. 5(a-d), respectively. The resulting GISAXS patterns depend on the size, shape, and arrangement of the nanostructured layer. The position of the direct beam stop (DBS), specular beam stop (SBS), and intermodular gap (IMG) are marked along with the position of the horizontal and vertical cuts in Fig. 5(a). The Yoneda peak position for Co and C60 is also marked in Fig. 5(c). The observed GISAXS pattern for sample S4 (shown in Fig. 5d) shows the regular intensity stripe features along the out-of-plane axis due to the parallel orientation of Co and $C_{60}$ layers to the substrate.

The line profiles of $I(q_y)$ and $I(q_z)$ extracted for qualitative analysis of the GISAXS data using the DPDAK software[37] are shown in Fig. 5(e) and (f), respectively. The cuts along the $q_y$ and $q_z$ directions contain information about the sample's lateral and vertical structural parameters. In Fig. 5(e), the broad hump at $q_y = 0.00875$ Å$^{-1}$, which is marked by dotted lines, does not shift on depositing the Co film. Thus, this can explicitly correspond to the large $C_{60}$ molecules with a lateral size ($2\pi/q_y$) of around 717.7 Å. Additionally, a small peak appears at larger $q_y$ values for samples S1-S4, which shifts to lower q values as thickness increases. This peak appears at $q_y$ value greater than 0.1 A for samples S1 and S2. This suggests that the interparticle distance between Co clusters is around 62.8 Å$^{-1}$. This can be understood as due to the high surface energy of Co, the Co clusters are formed, which gets diffused inside the $C_{60}$ film, leading to a huge distribution of interparticle distance. Thus, a broad peak appears at lower-thickness Co films. While in the case of sample S3, the position of the peak is at $q_y = 0.095$ Å$^{-1}$, which shifts towards lower $q_y = 0.012$ Å$^{-1}$ for sample S4. The positions of the peaks correspond to the lateral spacings of about 66.1 Å and 79.5 Å in samples S3 and S4, respectively. This change in peak position denotes an increase in the average center-to-center distance of the existing Co clusters caused by nearby clusters' coalescence.[38]

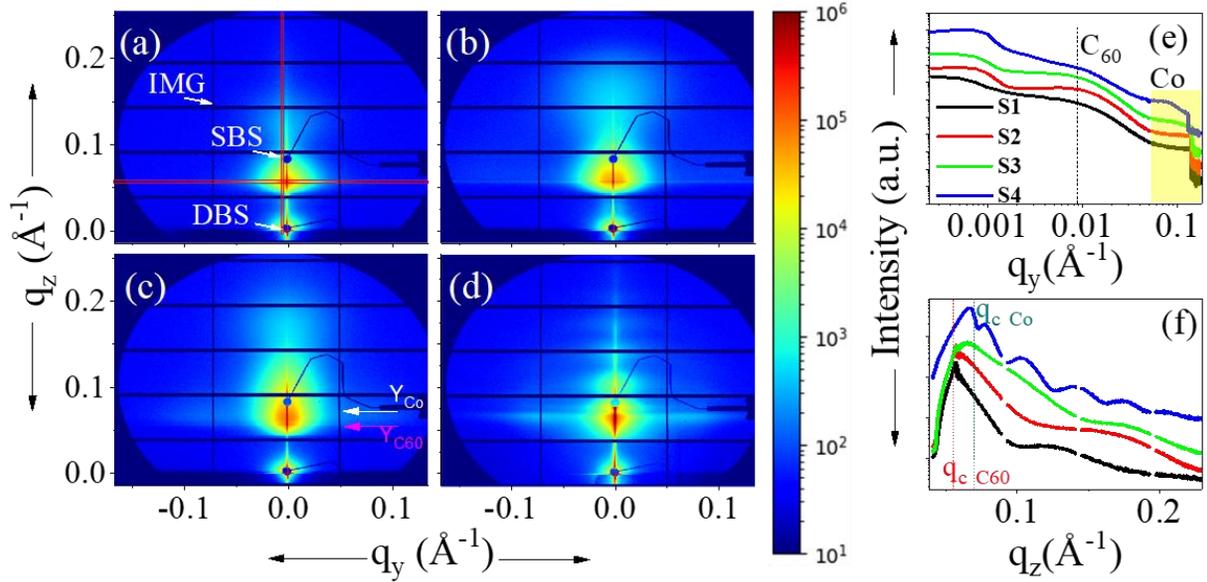

**Figure 5** (a)-(d) shows selected GISAXS scattering patterns of samples S1-S4. (e)-(f) extracted the $I(q_y)$ and $I(q_z)$ (Å$^{-1}$) from images for all the samples.

Another fascinating feature of the scattering profiles $I(q_z)$ is observed where the intensity shifts from lower to higher $q_z$ values, as shown in Fig. 5(f). The region known as the Yoneda peak is observed at the material-specific critical angle ($\alpha_c$). This angle is determined by the real part of the X-ray refractive index ($\delta$), where $\alpha_c = \sqrt{(2\delta)}$. Notably, $\delta$ is directly related to the average charge density of the material. The theoretically calculated values of the critical angle $\alpha_c$ for Co (density: 8.9 /cc) and $C_{60}$ (density: 1.65 g/cc) at 12.11 keV are 0.276° and 0.123°, respectively. Thus, the corresponding $q_c$ values for Cobalt and $C_{60}$ are 0.072 Å$^{-1}$ and 0.056 Å$^{-1}$ for an incident angle of 0.4°, also marked by dotted lines in Fig. 5(f). Thus, the shift in the peak position corresponds to the increased elemental charge density at the interface, leading to continuous film formation.

Figure 6(a) shows the GIWAXS image collected for sample S4 simultaneously with GISAXS measurements. The overall extracted integrated intensity profiles $I(q)$ in out-of-plane and in-plane geometry are plotted in Fig. 6 (b) and (c), respectively. For the quantitative analysis, the three peaks are fitted using a Gaussian function, and the obtained peak positions are 2.93 Å$^{-1}$, 3.11 Å$^{-1}$, and 3.30 Å$^{-1}$, which corresponds to hcp (100), (002) and (101) phases of Cobalt respectively. It is observed that there is a slight shift in the peak positions observed in the in-plane geometry, which suggests the presence of in-plane compressive stress in the Co film. Further, one can notice that the relative peak intensity shows a drastic change in both geometries. The normalized intensity $I_A$ for Co hcp (002) is 0.20, and 0.53 is calculated along out-of-plane and in-plane directions, conveying the presence of Co film's in-plane texturing along hcp (002). To further understand the preferential texturing direction in the Co plane, XRD analysis of sample S4 is conducted using two different geometries, keeping the momentum transfer vector (q ~ $q_{in}$ and $q_{out}$) in the plane (IP-XRD) and out of the plane (OP-XRD) of the film surface. These two geometries are depicted in Fig. 7 (7a and 7b).

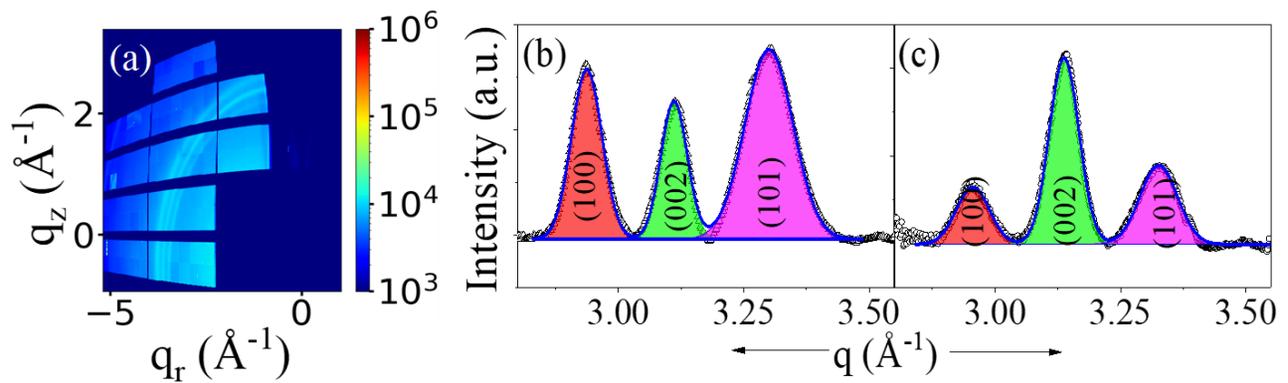

**Figure 6** (a) shows 2D GIWAXS images for sample S4, and (b) & (c) shows the extracted peak intensity profile integrated radially along out-of-plane and in-plane directions, respectively.

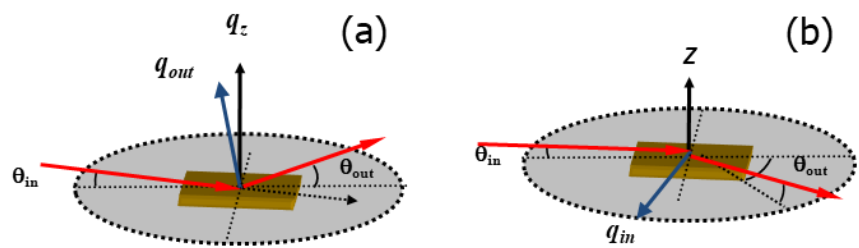

**Figure 7.** (a) and (b) Geometries used for the OP-XRD and IP-XRD measurements.

In the OP-XRD setup, the incident beam strikes the sample at a shallow angle ($\theta_{in} \sim 0.5°$) while the detector scans perpendicular to the film surface. This setup positions the $q_{out}$ in the film's normal direction, providing information about the scattering planes nearly parallel to the film surface. On the other hand, in the IP-XRD geometry, both the incident and diffracted beams form a grazing angle of 2.8° vertically, and the detector scans along the plane of the film. In this setup, the momentum transfer vector ($q_{in}$) is almost in the plane of the film, which preferentially provides information from the scattering planes perpendicular to the film surface. Further, the q vector is rotated in the azimuthal direction to investigate the preferential orientation of Co hcp (002) grains in the film plane.

Fitted XRD patterns in OP geometry are shown in Fig. 8(a), and IP XRD patterns with momentum transfer vector ($q_{in}$) along $\phi = 0°, 10°, 90°$ are shown in Fig. 8 (b-d), respectively. The OP-XRD pattern of the S4 film exhibits three distinct broad peaks corresponding to (100), (002), and (101) planes of HCP-Co, while in IP-XRD, texturing along HCP (002) peak is observed. In the in-plane direction, the relative peak intensity $I_A$ of HCP (002) is different when the direction of the $q_{in}$ within the sample plane is varied.

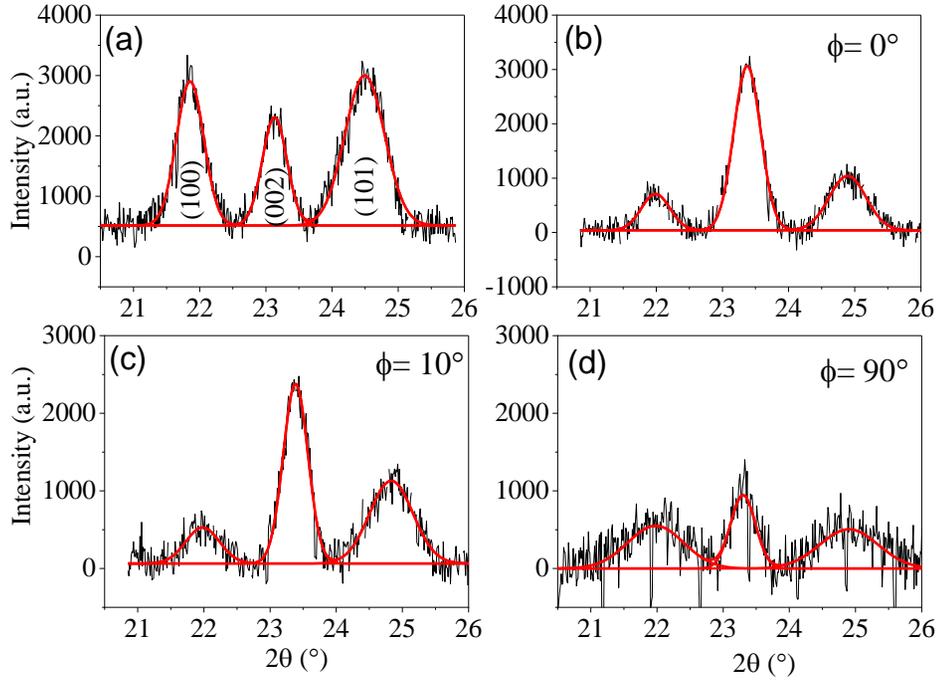

**Figure 8.** (a) Fitted XRD patterns in out-of-plane geometry and (b-c) IP-XRD data for the momentum transfer vector ($q_{in}$) at different azimuthal directions in the film plane.

To get qualitative information, position ($q$), normalized relative intensity ($I_A$), and $d$-spacings ($d$), corresponding to HCP (100), (002), and (101), are extracted after fitting the obtained XRD patterns and shown in Table 2. It is important to note that the d-spacings are lesser in the in-plane direction, suggesting the presence of in-plane compressive stress in Co film. Further, it can be seen from the table that the normalized relative intensity of HCP (002) is higher in the in-plane direction, suggesting the texturing of Co film along the HCP(002) phase. It is to be noted that Co film is textured in the in-plane direction, but the preferential orientation of the HCP(002) phase is observed along $\phi = 0°$.

**Table 2.** Fitting parameters obtained from XRD patterns taken along in-plane (IP) and out-of-plane (OP) geometry for S4; typical error bars obtained from the least square fitting of the XRD data approx. ± 5% in the value of the normalized area, $I_A$.

| Geometry | OP-XRD | IP- phi 0° | IP- phi 10° | IP- phi 90° |
|---|---|---|---|---|
| $I_{A(002)}$ | 0.211 | 0.59 | 0.48 | 0.27 |
| $d_{(002)}$ (Å) | 2.10 | 2.08 | 2.08 | 2.08 |

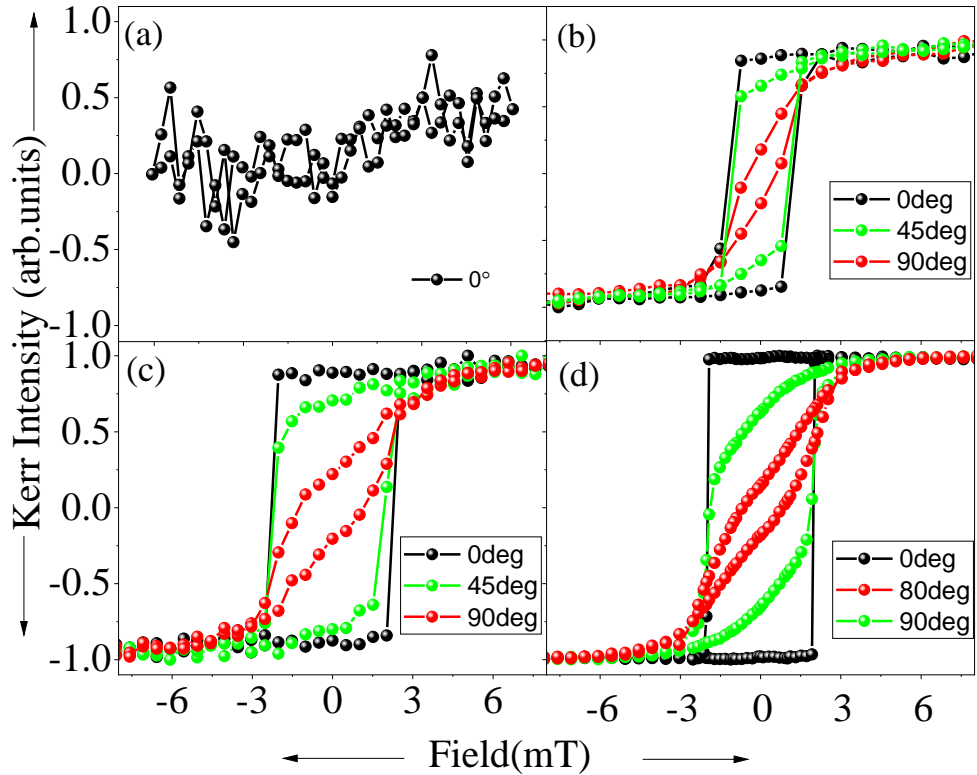

**Figure 9** (a)-(d) Hysteresis loops along different magnetization directions for samples S1 (Co ~10 Å), S2 (Co ~25 Å), S3 (Co ~40 Å) and S4 (Co ~180 Å).

The magnetic properties of the samples are investigated using MOKE hysteresis loops for samples S1-S4, as shown in Figure 9. No hysteresis loop is observed at lower thickness, i.e., in sample S1 (Figure 9 a). As the thickness of the Co layer increases in S2-S4, the hysteresis loop starts to manifest, and substantial variation in the shape of hysteresis with azimuthal angles is observed. Along the easy axis, the shape of the hysteresis is close to rectangular with a remanence ratio of Mr/Ms close to 1, which decreases with the increasing azimuthal angle. This drastic change in the shape of the hysteresis loops indicates the presence of magnetic anisotropy in the samples S2-S4. The magnetic properties of Co films in samples S2-S4 exhibit a distinct behaviour compared to those deposited on inorganic substrates. In the case of Silicon substrates, it is to be noted that no magnetic anisotropy was observed in reference samples S1-S3 (not shown here). The observed preferential orientation of the HCP (002) in the in-plane direction suggests that the anisotropy in Co film is magneto-crystalline in nature. The variation of the coercivity with azimuthal angle is shown in Fig. 10 for samples S4 and S4 ref. Compared to the reference sample, sample S4 exhibits higher coercivity values and enhanced magnetic anisotropy in the Co film, attributed to the Co-$C_{60}$ interface. The presence of local maxima is also seen at the nominal hard axis (i.e., 90°) other than 0° and 180° in the coercivity variation for sample S4 compared to the reference sample.

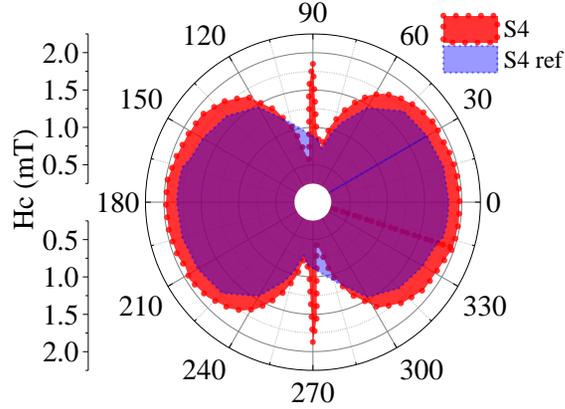

**Figure 10** Angle dependent coercivity ($H_C$) plots for samples S4 and S4 ref to compare the anisotropy symmetry.

The azimuthal angular dependence of the $M_R/M_S$ for samples S2-S4 is plotted in Fig. 5(a-c) to understand the magnetic anisotropy and magnetization reversal mechanism in the samples. In sample S2, the variation aligns with the anticipated uniaxial anisotropy characteristics, following a $|Cos(\theta)|$ variation. This corresponds to the geometric projection of the magnetization of the easy axis onto the field axis. However, in samples S3 and S4, an anomaly is observed along the hard axis and its immediate surroundings, in addition to the cosine variation. It is seen in Fig.11(b) and (c) that the Mr/Ms variation displays a distinct and prominent peak centered around the nominal hard axis.

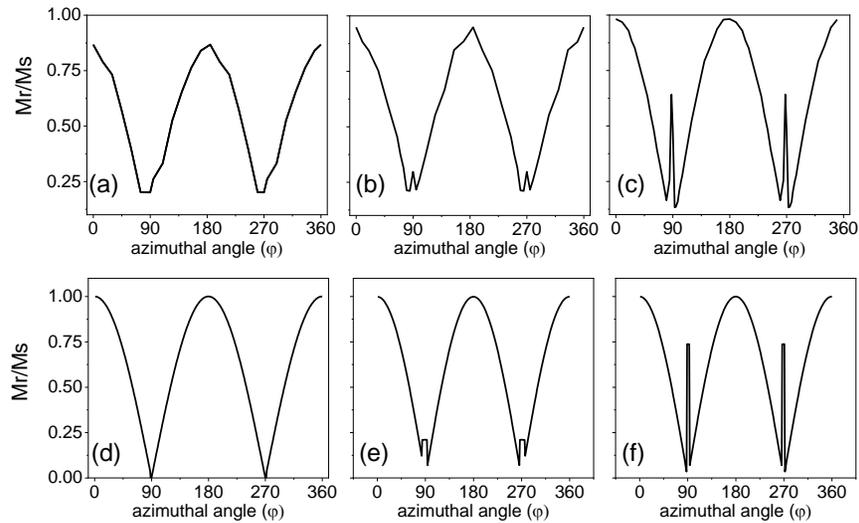

**Figure 11** (a-c) shows observed and (d-f) shows simulated Azimuthal angle dependence of remanence for samples S2, S3, and S4 respectively.

This behavior significantly diverges from the ideal Stoner-Wohlfarth single-domain particles, in which the hard axis exhibits zero remanence and coercivity as magnetization reversal is governed by coherent rotation. A similar kind of anomaly, i.e., high values of the remanence and the coercivity in the hard axis, is found in polycrystalline samples with in-plane uniaxial magnetic anisotropy.[39] This non-ideal hard axis behaviour has also been understood in terms of domain wall interactions.[40] In the present case, this unusual variation in the remanence is understood by the inhomogeneous texturing along the HCP (002) in the film plane (as discussed in the previous section). This leads to the deviation in the local anisotropy axis; hence, magnetization is no longer homogeneous in the film plane.

To gain deep insight into how the deviation in local anisotropy causes non-ideal behaviour in $H_C$ and $M_R/M_S$ variation along the hard axis, a two-grain Stoner Wohlfarth model approach was utilized with intergranular exchange coupling and misalignment in the grain axis.[39,41] For this, we minimize the energy of the system with two different grains having anisotropy energy, Zeeman energy, and exchange energy. The energy of the system is given as

$$E = J\cos(\theta_1 - \theta_2) + H\left[\cos(\theta_1 - \varphi) + \cos(\theta_2 - \varphi)\right] + \frac{1}{2}K[\sin^2(\theta_1 - \alpha/2) + \sin^2(\theta_2 + \alpha/2)]$$

where, J represents the constant for intergranular exchange coupling between two grains in the Co film, H is the applied magnetic field, K is the uniaxial magneto-crystalline anisotropy constant, $\alpha$ is the misalignment between two grains/easy axes (EA), $\varphi$ is the azimuthal angle between the nominal easy axis and applied magnetic field, $\theta_1$, and $\theta_2$ are the angle that magnetization vector $\mathbf{M_{S1}}$ and $\mathbf{M_{S2}}$ makes with nominal easy axis ($\varphi=0°$). The Mr/Ms variation is then simulated in the two-grain system for J/K= 0.35, $\alpha$= 0°, J/K =0.35, $\alpha$= 10° and J/K = 0.48, $\alpha$= 10° as shown in Fig. 11(d-f) respectively.

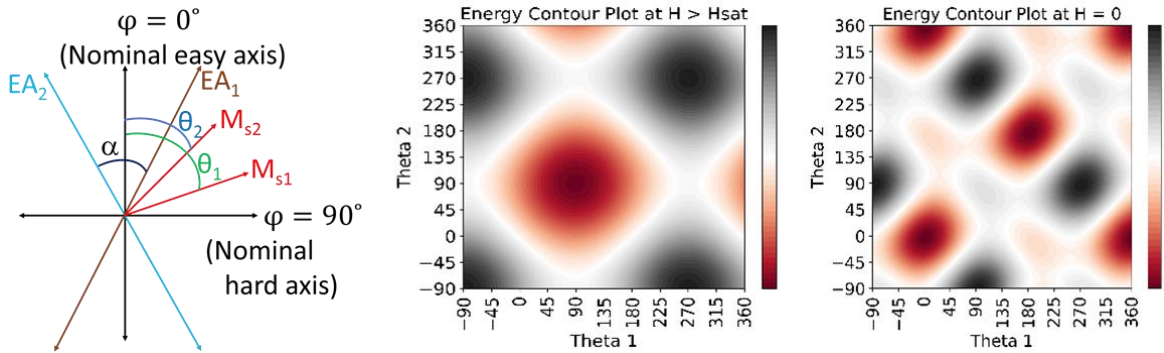

**Figure 12** (a) Rotation of magnetization vectors in a two-grain system upon application of a magnetic Field. (b)&(c) Energy contour plots with varying $\theta_1$ and $\theta_2$ for applied field H> $H_{sat}$ and H=0 respectively.

The schematic of the rotation of magnetization vectors in the two-grain system upon applying a magnetic field is represented in Fig. 12(a). When the system is in a demagnetized state, in the absence of exchange coupling between the two grains, $\theta_1$ and $\theta_2$ would be $\pm\alpha/2$ respectively, on including exchange interactions, $\theta_1 - \theta_2 < \alpha$ i.e., $0 < \theta_1 < \alpha/2$ and $-\alpha/2 < \theta_2 < 0$. This can be understood by considering the case when the H/K= 10, J/K=0.35 and $\alpha$=10°. Thus, when applying the magnetic field towards the easy axis of magnetization, the magnetization vectors $M_{S1}$ and $M_{S2}$ align with the field direction, thereby minimizing the energy of the system. After removal of the applied field, magnetization vectors switch to the nearest minima in the energy landscape, i.e. $(\theta_1, \theta_2) \sim (2.94°, -2.94°)$. While if the applied magnetic field (H>$H_{sat}$) is along the nominal hard axis, i.e., $\varphi$ =90°, energy minimum/point of stability occurs around $(\theta_1, \theta_2) \sim (89.48°, 90.51°)$ as shown in Fig. 12(b). After removing the field, a local minimum appears in the energy contour plots for $\theta_1 =15.68°$ and $\theta_2 =164.31°$ as shown in Fig.12(c). To bring the system from the nearest local minima to point of stability, additional energy needs to be provided to the system, which gives rise to high coercive field values along the nominal hard axis. Further, remanent magnetization in a sample can be understood as the projection of the magnetization onto the field axis after removing the applied field. Thus, the observed unusual remanence behavior along the hard axis can be understood due to the appearance of local minima for $\theta_1 =15.64°$ and $\theta_2 =164.38°$ which leads magnetization vectors $\mathbf{M_{S1}}$ and $\mathbf{M_{S2}}$ align in such a way that Mr/Ms would become Cos (90-15.64) + Cos (90-164.38) and is not equal to 0. Thus, instead of observing minima at the $\varphi =90°$, a sharp peak is observed at the nominal hard axis, as shown in Fig.11(b-c).

It is important to note that no peak is observed at the hard axis up to a critical misalignment angle and maximum intergranular exchange coupling. For the values of $\alpha_c <10°$ and J/K >0.5, the exchange coupling energy is so dominant that it compensates for the misalignment of the anisotropy axes such that both $\mathbf{M_{S1}}$ and $\mathbf{M_{S2}}$ display the same rotation during magnetization reversal. In that case, the system resembles the behavior of the single grain in the Stoner Wohlfarth model. When the misalignment angle increases from 10°, the anomalous peak starts to develop and becomes broader upon increasing $\alpha$. For J/K= 0.35, on varying the misalignment angle from 10° to 20°, the local minimum appears at (27.48°, 152.51°), leading to the broadening in the anomalous peak since when misalignment is 20°, magnetization vectors do not trace the initial points of minima even when the field is applied along $\varphi =80°$. Likewise, when we increase the J/K ratio from 0.35 to 0.45 while keeping the misalignment angle fixed at 10°, we observe the minimum point occurring at (31.98°, 148.01°). Thus, the peak observed in the simulated plots is sensitive to the J/K ratio and misalignment angle $\alpha$. It is the modifications in the position of local minima and the alignment of magnetization vectors (i.e., for such combination of ($\theta_1$, $\theta_2$)) due to which the cosine component of magnetization increases along the nominal hard axis. In the present study, the misalignment in the two grains is observed in samples S3 and S4. Additionally, an increase in the intergranular exchange coupling constant has also been observed in increasing the thickness of Co film. The discrepancy between the simulated and experimental data arises because the theoretically simulated remanence variation is characteristic of a two-grain system. However, in samples S3 and S4, the polycrystalline Co film comprises multiple grains with varying misalignment angles.

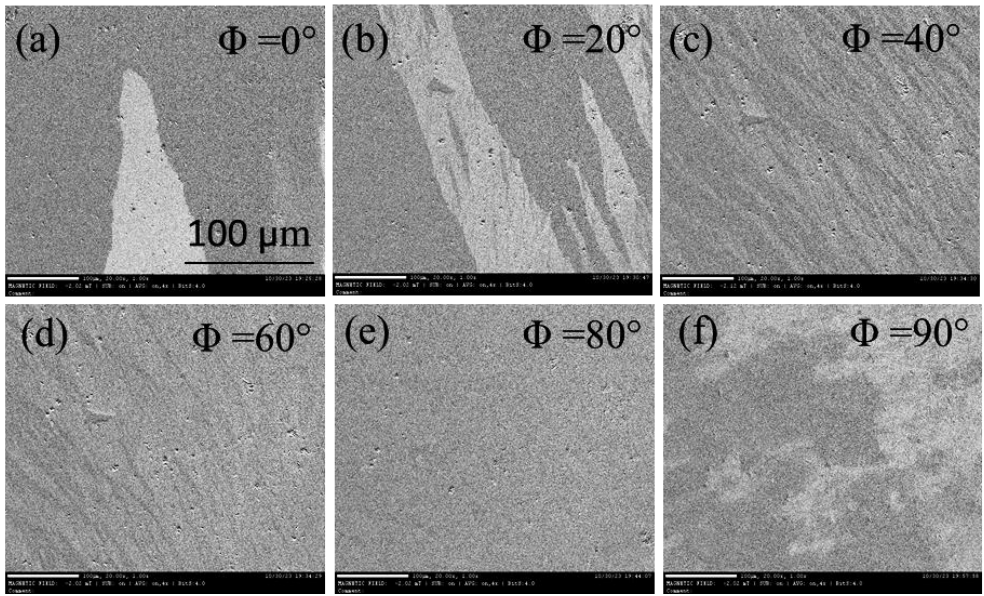

**Figure 13**. Domain images near the coercivity for samples S4 are shown in (a)–(f), respectively.

The observed anomaly in the coercivity and remanence variation along the hard axis also corroborates with the domain images taken near the coercive field simultaneously with the hysteresis loop measurements using Kerr microscopy. Figure 13 shows the captured domain images near the coercive field at different azimuthal angles for sample S4. The domain images are measured using magneto-optic Kerr effect (MOKE) based microscopy at room temperature in longitudinal mode by varying the angle ($\phi$) between the easy axis and the applied field direction. The scale bars of the images for sample S4 are shown in Fig. 13 (a). The applied field direction shown in image (a) was kept constant during all the measurements, and the sample was rotated accordingly.

Figure 13 (a-f) shows the domain images corresponding to sample S4. It can be seen along the easy axis ($\phi=0°$) that large branch domains are observed. As the azimuthal angle is rotated away from the easy axis, the domain size shrinks, and stripe domains become apparent.[42] Near the hard axis ($\phi=80°$), no domains were observed as the magnetization reversal mechanism is taking place via magnetic moment rotation, while at the nominal hard axis ($\phi=90°$), the non-uniform spatial distribution of magnetization is observed. The domain images obtained from Kerr microscopy measurements indicate the presence of an unusual, non-uniform state of magnetization along the nominal hard axis. Thus, the observed Kerr microscopy images further confirms the existence of anomalous magnetization behaviour at the nominal hard axis in the remanence and coercivity variation.

## CONCLUSIONS

The present study illustrates the growth-induced evolution and correlation in the structural and magnetic properties of the Co film on the $C_{60}$ layer. Due to the mechanical softness of $C_{60}$, ferromagnetic Co atoms penetrate the $C_{60}$ film near the interface. The formation of diffused small Co clusters inside the $C_{60}$ film (near interface) is found to be responsible for the disappearance of ferromagnetism in Co film at a thickness of less than 10 Å due to their paramagnetic nature. Unlike inorganic substrates, magnetic anisotropy begins to develop at a deposition thickness of Co around 25 Å on the $C_{60}$ layer, which might be due to the non-planar structure of the $C_{60}$ buckyballs. At higher thicknesses, the observed magnetic anisotropy differs from conventional systems such as Co, Fe, and Ni on Si and $SiO_2$ substrates due to the anomalies in remanence variation, magnetic domains, and higher coercivity near the hard axis of magnetization. These anomalies are attributed to the distribution of magneto-crystalline anisotropy and inhomogeneous strain at the Co/$C_{60}$ interface, as supported by domain imaging showing a non-uniform spatial distribution of magnetic domains. Unusual magnetic anisotropy is understood using a theoretical model based on a two-grain Stoner-Wohlfarth model, which considers varying misalignment angles and exchange coupling constants between multiple magnetic grains. This misalignment and coupling between the grains is attributed to the dispersion in the anisotropy axis caused by the formation of higher interface roughness due to Co film penetration into the $C_{60}$ layer. The current study investigates the role of morphological and structural variations in the observed magnetic anomalies, which in turn can be tailored to get the desired functionality of Co and $C_{60}$-based organic spin valve devices. Gaining a comprehensive understanding of magnetic anisotropy in metal-organic thin films and developing effective control over it will empower researchers to tailor film properties and optimize their functionality.

## ACKNOWLEDGEMENT

Financial support from the Department of Science and Technology, Government of India (project CRG/2021/003094) and travel support within the framework of the India@DESY collaboration is gratefully acknowledged. Parts of this research were carried out at the P03 beamline of PETRA III at DESY, a member of the Helmholtz Association HGF. Thanks to Mr. Mohan Gangrade for the AFM measurements. Authors also acknowledge Dr. Mukul Gupta and Ms. Akshaya A for Co and Al deposition.